\documentclass[twoside,twocolumn]{article}
\usepackage[siamtitle,nocopyright,smallbib,toppagenumbers]{jeffeproc}
\usepackage{jeffe}
\usepackage{epsfig}
\usepackage{myconcrete,euler}
\usepackage{url}

\newtheorem{lemma}{Lemma}[section]
\newtheorem{theorem}[lemma]{Theorem}
\newtheorem{corollary}[lemma]{Corollary}

\def\SVM{{Vis(\Delta\mid\nobreak P)}}
\def\trap{\tau}

\begin{document}

\pagestyle{myheadings}
\markboth{Finite-Resolution Hidden Surface Removal}
         {Jeff Erickson}

\urldef{\myurl}\url{http://www.uiuc.edu/~jeffe}
\urldef{\myemail}\url{jeffe@cs.uiuc.edu}

\urldef{\paperurl}\url{http://www.uiuc.edu/~jeffe/pubs/gridvis.html}

\title{Finite-Resolution Hidden Surface Removal\thanks{
	Research partially supported by National Science Foundation
	grant DMS-9627683, U.S. Army Research Office MURI grant
	DAAH04-96-1-0013, by a Sloan Fellowship.  See \paperurl\ for
	the most recent version of this paper.}}

\author{Jeff Erickson\thanks{
	Department of Computer Science, University of Illinois,
	Urbana-Champaign; \myurl; \myemail.  Portions of this research
	were done at the Center for Geometric Computing, Duke
	University.}}

\date{}

\maketitle

\begin{abstract}
\small We propose a hybrid image-space/object-space solution to the
classical hidden surface removal problem: Given $n$ disjoint triangles
in $\Real^3$ and $p$ sample points (``pixels'') in the $xy$-plane,
determine the first triangle directly behind each pixel.  Our
algorithm constructs the \emph{sampled visibility map} of the
triangles with respect to the pixels, which is the subset of the
trapezoids in a trapezoidal decomposition of the analytic visibility
map that contain at least one pixel.  The sampled visibility map
adapts to local changes in image complexity, and its complexity is
bounded both by the number of pixels and by the complexity of the
analytic visibility map.  Our algorithm runs in time $O(n^{1+\e} +
n^{2/3+\e}t^{2/3} + p)$, where $t$ is the output size.  This is nearly
optimal in the worst case and compares favorably with the best
output-sensitive algorithms for both ray casting and analytic hidden
surface removal.  In the special case where the pixels form a regular
grid, a sweepline variant of our algorithm runs in time $O(n^{1+\e} +
n^{2/3+\e}t^{2/3} + t\log p)$, which is usually sublinear in the
number of pixels.\par
\end{abstract}

\section{Introduction}

Hidden surface removal is one of the oldest and most important
problems in computer graphics.  Informally, the problem is to compute
the portions of a given collection of geometric objects, typically
composed of triangles, that are visible from a given camera position
and orientation in $\Real^3$.  In order to simplify calculation (and
explanation), a projective transformation is applied so that the
camera is at $-\infty$ on the $z$-axis and all vertices have positive
$z$-coordinates, so that the desired image is the orthographic
projection of the objects onto the $xy$-plane.  We will follow the
computer graphics convention that the $y$-axis is vertical, the
$x$-~and $z$-axes are horizontal, and the positive $z$-axis points
into the image, directly away from the camera.

Historically, there are two different approaches to solving the hidden
surface removal problem: \emph{object space} and \emph{image
space}~\cite{sss-cthsa-74}.  Object-space (or \emph{analytic}) hidden
surface removal algorithms compute which object is visible at every
point in the image plane.  Image-space algorithms, on the other hand,
compute only the object visible at a finite number of sample points.
We will refer to the sample points themselves as ``pixels'', since
usually there is one sample point per pixel in the final
finite-resolution output image.  (Image-space algorithms that compute
sub-pixel features do so by sampling a small constant number of points
within each pixel area~\cite{fdfh-cgpp-90}.)

The output of an object-space hidden surface removal algorithm is the
projection of the forward envelope\footnote{This would be called the
``lower envelope'' if the $z$-axis were vertical.} of the objects onto
the image plane.  The resulting planar decomposition is called the
\emph{visibility map} of the objects.  Each face of the visibility map
is a maximal connected region in which a particular triangle, or no
triangle, is visible.  McKenna~\cite{m-wcohs-87} described the first
algorithm to compute visibility maps in $\Theta(n^2)$ time, where $n$
is the number of input triangles; see also~\cite{d-qbhle-86}.  This is
optimal in the worst-case.  Unfortunately, McKenna's algorithm
\emph{always} uses $\Theta(n^2)$ time and space, even when the
visibility map is much simpler.  This shortcoming led to the
development of several \emph{output-sensitive} algorithms, whose
running time depends not only on $n$, the number of triangles, but
also on $v$, the number of vertices of the visibility map.  The
fastest algorithm currently known, an improvement by Agarwal and
\Matousek~\cite{am-rsps-93} of an algorithm of de~Berg
\etal~\cite{bhosk-ershs-94}, runs in time $O(n^{1+\e} +
n^{2/3+\e}v^{2/3})$.  For more details on these and other object-space
algorithms, see the comprehensive survey by Dorward~\cite{d-soshs-94}.

The primary disadvantage of the object-space approach is the
potentially high complexity of the visibility map, which may be much
larger than the number of pixels in the desired output image, even for
reasonable input sizes.  Even when the visibility map is not overly
complex, it may contain features that are significantly smaller than
the area of a pixel and thus do not contribute to the final image.
This is especially problematic for applications of hidden-surface
removal such as form-factor calculation, where the desired output
image may have very low resolution \cite{sp-rgi-94}.

For image-space algorithms, on the other hand, the ultimate goal is to
compute, for each pixel in the finite-resolution output image, which
triangle is visible at that pixel.  The most common image-space
approach is the \emph{$z$-buffer} algorithm introduced by Catmull
\cite{c-sacdc-74}.  This algorithm loops through the triangles,
determining the pixels that each triangle covers in the image plane;
each pixel maintains the smallest $z$-coordinate of any triangle
covering that pixel.  While this algorithm can be implemented cheaply
in hardware, it can still be quite slow when the number of triangles
and number of pixels are both large.

Another common image-space approach is \emph{ray casting} (also known
as \emph{ray tracing} and \emph{ray shooting}): Shoot a ray from each
pixel in the positive $z$-direction and compute the first triangle it
hits.  Using using the best known unidirectional ray-shooting data
structure, due to Agarwal and Sharir~\cite{as-anspt-93}, we obtain an
algorithm with running time $O((n + n^{2/3}p^{2/3} + p)\log^3 n)$,
where $n$ is the number of triangles and $p$ is the number of pixels.
Erickson's lower bound for Hopcroft's problem~\cite{e-nlbhp-96}
suggests that this algorithm is close to optimal in the worst case,
even for the simpler problem of deciding whether any ray hits a
triangle.  In practice, ray-shooting queries are answered by walking
through a decomposition of space determined by the triangles, such as
an octtree~\cite{g-ssfrt-84}, triangulation~\cite{af-acrsm-97}, or
binary space partition \cite{m-ehsrt-98,py-ebsph-90}.  See
\cite{af-acrsm-97,hs-parss-95} for related theoretical results.

Neither $z$-buffers nor ray casting exploit \emph{spatial coherence}
in the image.  If the visible triangles are fairly large, then the
same triangle is likely to be visible through several pixels; however,
both algorithms compute the triangle behind each pixel independently.
Spatial coherence is exploited to some extent by more complex
techniques such as Warnock's subdivision algorithm~\cite{w-hsacg-69},
hierarchical $z$-buffers~\cite{gkm-hzbv-93}, hierarchical coverage
masks~\cite{g-hptcm-96}, and frustum casting~\cite{ta-fcpir-98}, which
construct a recursive quadtree-like decomposition of the image.
However, this decomposition can be much more complex than the
visibility map if, for example, the image contains several long
diagonal lines.  In particular, if the pixels lie in a regular
$\sqrt{p}\times\sqrt{p}$ grid, the decomposition can have complexity
$\Theta(v\sqrt{p})$.

A few hidden surface removal algorithms work simultaneously in both
image and object space \cite{hh-btpo-84,wa-hsrup-77}.  The basic idea
for these algorithms is to traverse the objects in order from front to
back (\ie, by increasing ``distance'' from the camera), decomposing
the image plane using the boundaries of the objects and reverting to
ray casting when any region of the image plane contains only a single
pixel.  Of course, there are sets of triangles do not have a
consistent depth order, and these algorithms will produce incorrect
output if such as set is given as input.  While a depth order can
always be guaranteed by first decomposing the triangles with a
binary-space partition tree, this could produce $\Theta(n^2)$ triangle
fragments in the worst case~\cite{py-ebsph-90}.  One exception to the
depth-order requirement is Weiler and Atherton's
algorithm~\cite{wa-hsrup-77}, which decomposes the image plane into
regions within which the triangles can be depth-ordered; this
algorithm can also produce a quadratic number of fragments.  The image
decompositions produced by these algorithms produce cannot be analyzed
either in terms of the complexity of the visibility map, since they
can decompose triangles even when all depth cycles are invisible, or
in terms of the number of pixels, since they can produce many
fragments that do not contain a pixel at all.

\medskip In this paper, we propose another hybrid approach to hidden
surface removal that exploits both spatial coherence and finite
precision.  In Section~\ref{S:def}, we define the \emph{sampled
visibility map} of a set of triangles with respect to a set of pixels.
Like other image-decomposition schemes, the sampled visibility map
adapts to local changes in the image complexity, but unlike previous
approaches its complexity is easily bounded both by the complexity of
the analytic visibility map and by the number of pixels.

We describe an output-sensitive algorithm to construct the sampled
visibility map in Section~\ref{S:alltraps}.  Our algorithm runs in
time ${O(n^{1+\e} + n^{2/3+\e}t^{2/3} + p)}$, where $t$ is the number
of trapezoids in the output.  This matches the performance of Agarwal
and \Matousek's visibility map algorithm when $t=\Theta(v)$, and
almost matches Agarwal and Sharir's ray-casting algorithm when
$t=\Theta(p)$.  Our algorithm does not require the triangles to have a
consistent depth order, nor does it decompose the triangles into
orderable fragments.  A variant of our algorithm allows a sequence of
pixels to be specified online, at an additional amortized cost of
$O(\log t)$ time per pixel.

The algorithms presented in Section~\ref{S:alltraps} assume that the
pixels are just arbitrary points in the $xy$-plane.  In
Section~\ref{S:grid}, we describe a faster algorithm for the common
special case where the pixels are the vertices of a rectangular grid.
The running time of our improved algorithm is $O(n^{1+\e} +
n^{2/3+\e}t^{2/3} + t\log p)$, which is sublinear in the number of
pixels unless the output is very large.

Finally, in Section \ref{S:outro}, we discuss some other applications
of our techniques and suggest directions for further research.

\section{Definitions}
\label{S:def}

Let $\Delta$ be a set of $n$ disjoint triangles in $\Real^3$, where
every vertex has positive $z$-coordinate.  We say that a triangle
$\triangle\in\Delta$ is \emph{visible} at a point $\pi$ in the
$xy$-plane if a ray from $\pi$ in the positive $z$-direction hits
$\triangle$ before any other triangle in $\Delta$.  The
\emph{visibility map} $Vis(\Delta)$ is a planar straight-line graph,
each face of which is a maximal connected region in which a particular
triangle in $\Delta$, or no triangle, is visible.  See Figure~1(a).
Let $v$ denote the number of vertices of $Vis(\Delta)$.

The \emph{trapezoidal decomposition} of $Vis(\Delta)$, denoted
$Trap(Vis(\Delta))$, is obtained by decomposing each face into
(possibly degenerate) trapezoids, two of whose edges are vertical
(\ie, parallel to the $y$-axis).  The vertical edges are defined by
casting segments up and/or down from each vertex into the face,
stopping when the segment reaches another edge of the face.  Faces are
decomposed individually, so only one vertical edge is added at a
``\textsf{T}'' vertex where one visible edge appears to overlap
another.  See Figure~1(b).

Finally, let $P$ be a set of $p$ points in the $xy$-plane, called
``pixels''.  The \emph{sampled visibility map} of $\Delta$ with
respect to $P$, denoted $\SVM$, is the subset of trapezoids in
$Trap(Vis(\Delta))$ that contain at least one pixel in $P$.  See
Figure~1(d).  Let $t$ denote the number of trapezoids in $\SVM$.
Clearly $t\le p$, since every trapezoid in $\SVM$ contains at least
one pixel.  Moreover, since $Trap(Vis(\Delta))$ contains at most $2v$
trapezoids, ${t\le 2v}$.

\begin{figure*}
\centerline{\footnotesize
\begin{tabular}{c@{\qquad\qquad}c}
     \epsfig{file=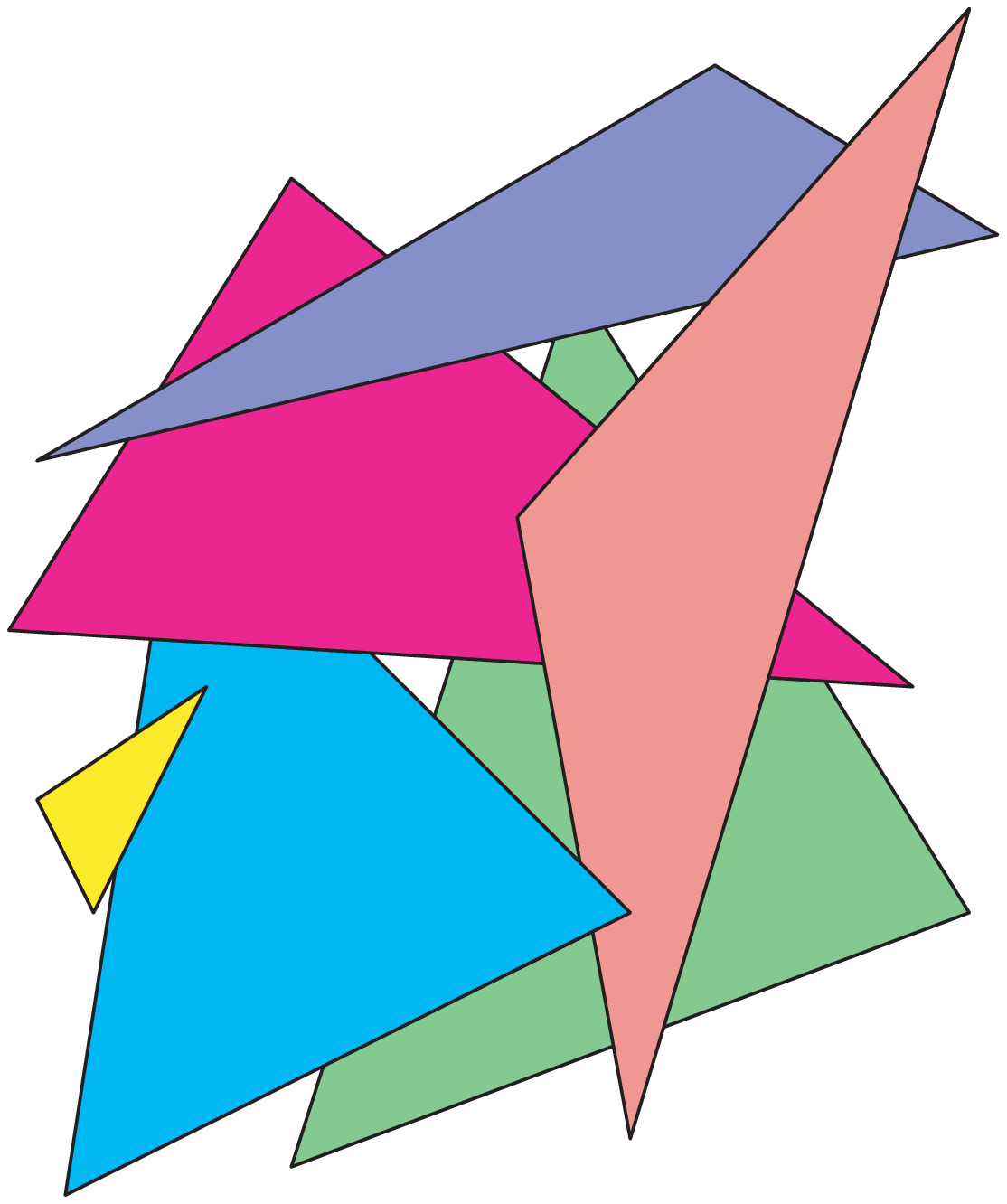,width=2.25in} &
     \epsfig{file=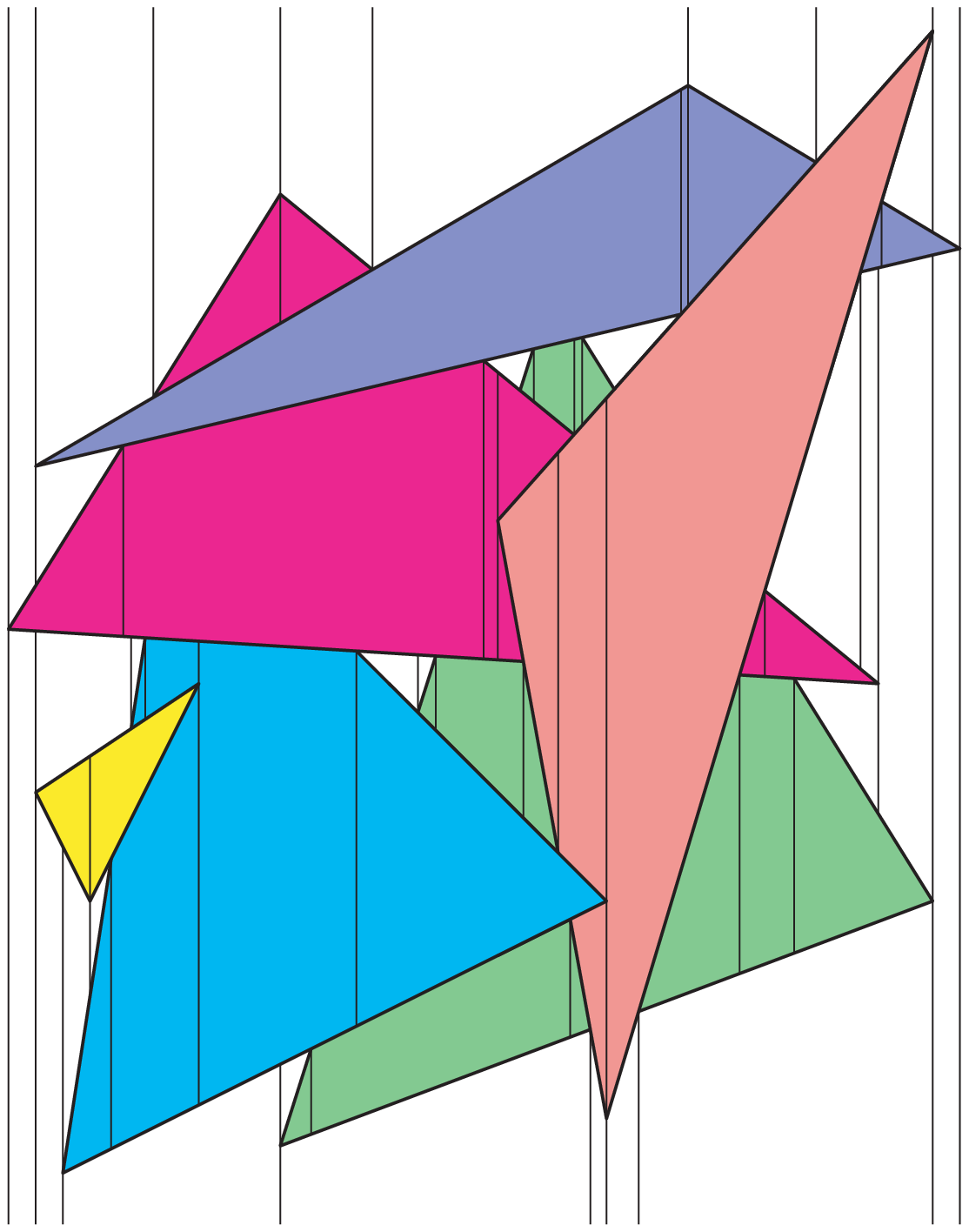,width=2.25in}
     \\
     (a) & (b)
     \\[2ex]
     \epsfig{file=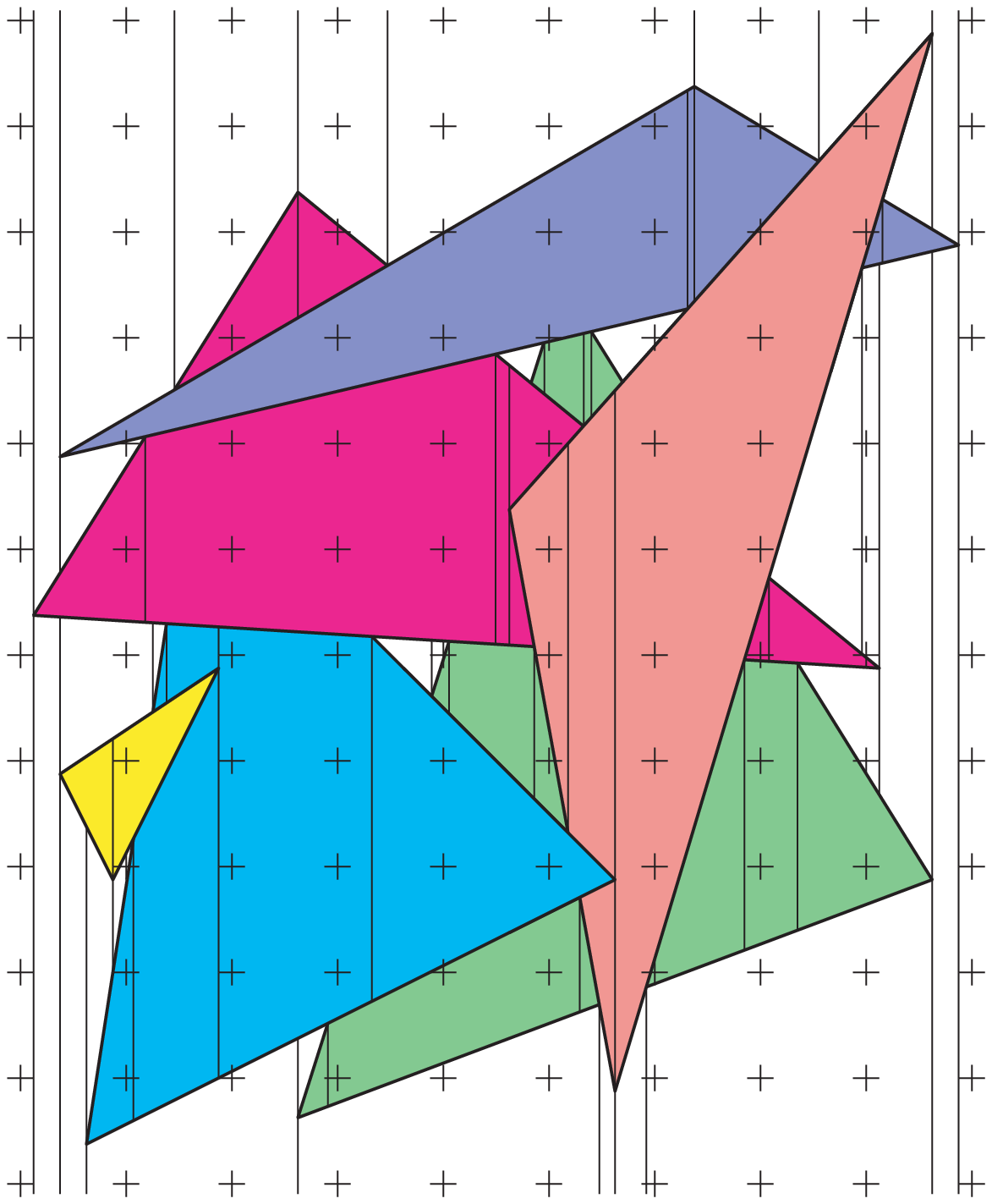,width=2.25in} &
     \epsfig{file=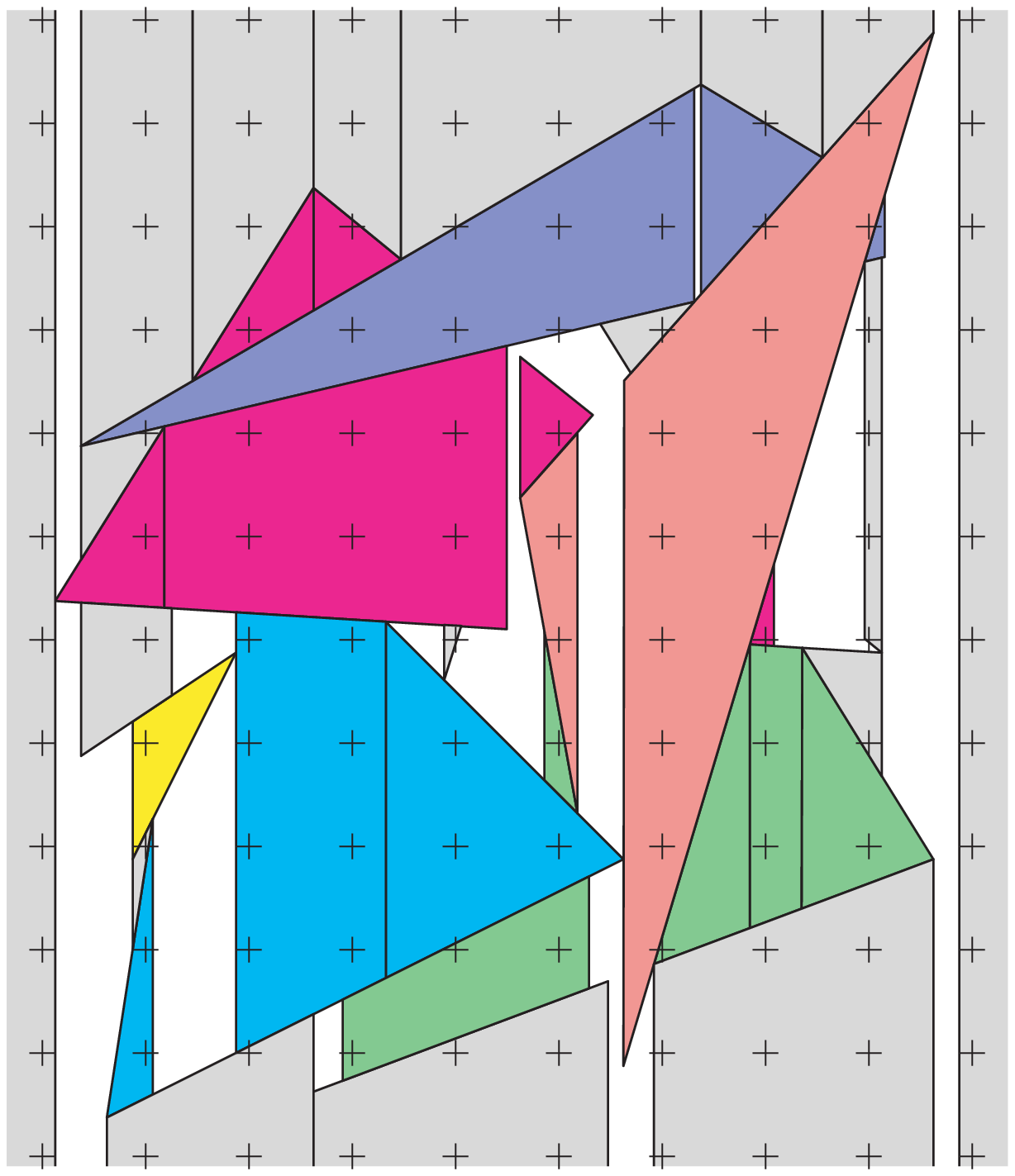,width=2.25in}
     \\
     (c) & (d)
\end{tabular}
}
\caption{(a)~The visibility map $Vis(\Delta)$ of a set $\Delta$ of
triangles, (b)~its trapezoidal decomposition $Trap(Vis(\Delta))$,
(c)~with a grid of pixels $P$, and (d)~the resulting sampled
visibility map $\SVM$.}
\end{figure*}

%
%
\begin{figure*}
\centerline{\footnotesize
\begin{tabular}{cccc}
	\epsfig{file=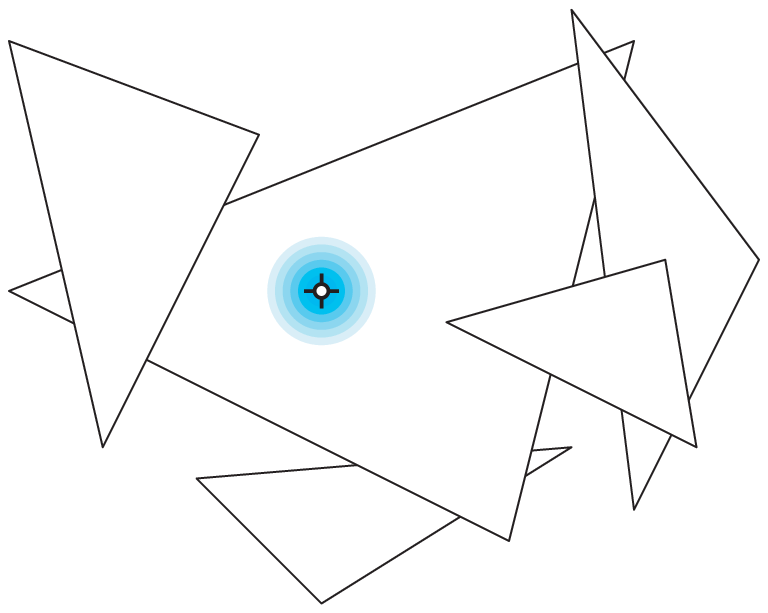,width=1.5in} &
	\epsfig{file=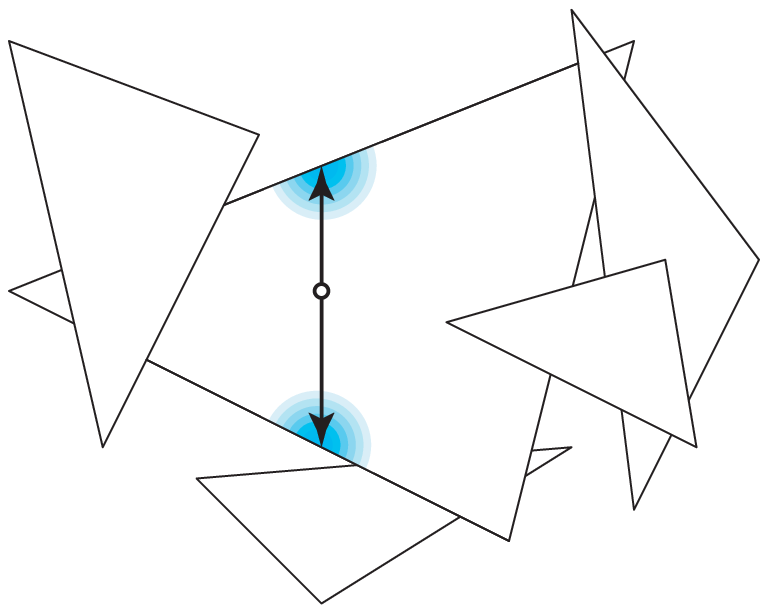,width=1.5in} &
 	\epsfig{file=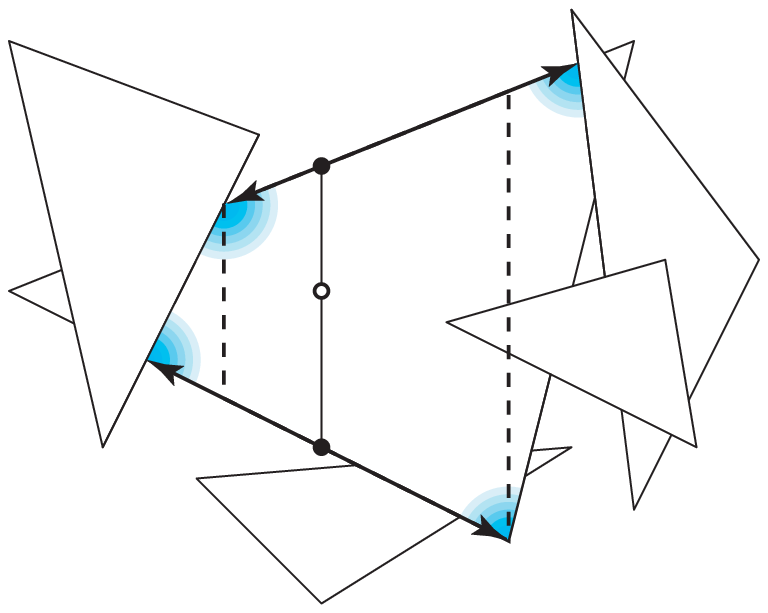,width=1.5in} &
	\epsfig{file=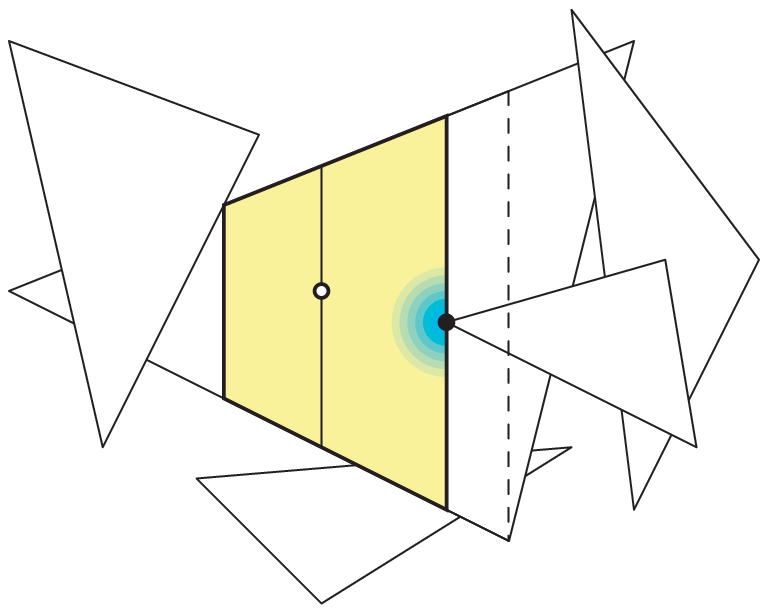,width=1.5in}
    \\
    (a) & (b) & (c) & (d)
\end{tabular}
}
\caption{Building one trapezoid in $\SVM$.  (a) Shoot a ray into the
scene through the pixel to the first triangle.  (b) Drag rays up and
down to find the top and bottom edges.  (c) Drag rays along the top
and bottom edges to find their (potential) endpoints.  (d) Narrow the
trapezoid by locating the nearest visible vertices on either side.}
\end{figure*}

\section{Building One Trapezoid in \boldmath$\SVM$}
\label{S:onetrap}

A \naive\ algorithm for constructing the sampled visibility map would
start by constructing $Vis(\Delta)$.  While this approach leads to an
algorithm that is nearly optimal in the worst case, it cannot give an
output-sensitive algorithm.  To obtain output-sensitivity, we
construct $\SVM$ one trapezoid at a time.  Specifically, for each
pixel $\pi\in P$, if it is unmarked, we determine the trapezoid
$\trap_\pi \in Trap(Vis(\Delta))$ that contains it and then mark all the
pixels contained in $\trap_\pi$.  We construct each trapezoid in four
stages, which are illustrated in Figure 2.

\bigskip\noindent\textbf{Stage 1.~Forward Ray Shooting.}
The first stage in constructing the trapezoid $\trap_\pi$ is to
determine the triangle visible at $\pi$; see Figure 2(a).  This is
done by answering a unidirectional ray-shooting query, exactly as in
the standard ray-casting algorithm.  Agarwal and
Sharir~\cite{as-anspt-93} describe a data structure that can answer
such queries in time $O((n/\sqrt{s})\log^3 n)$ using a data structure
of size $O(s\log^2 n)$, where $s$ can be chosen anywhere between $n$
and $n^2$.  The preprocessing time needed to construct this data
structure is $O(s\log^3 n)$.

Agarwal and Sharir's data structure is actually designed to answer
point stabbing queries for a set of triangles in the plane---How many
triangles contain the query point?  Like most geometric range
searching structures, their data structure defines a number of
\emph{canonical subsets} of the set of triangles.  For any point
$\pi$, the set of triangles that contain $\pi$ can be expressed as the
disjoint union of $O((n/\sqrt{s})\log^3 n)$ canonical subsets; in
particular, this implies that the triangles in any canonical subset
have a common intersection.  Their data structure stores the size of
each canonical subset, and a stabbing query is answered by summing up
the sizes of the relevant canonical subsets.  To obtain a
unidirectional ray-shooting data structure for our three-dimensional
triangles $\Delta$, it suffices to build Agarwal and Sharir's
point-stabbing structure for the $xy$-projection of $\Delta$.  Now the
triangles in any canonical subset have a consistent front-to-back
ordering, and the triangle visible through $\pi$ can be computed by
comparing the front-most triangles in the relevant canonical subsets.

\bigskip\noindent\textbf{Stage 2.~Vertical Ray Dragging.}
The second stage in our algorithm finds the top and bottom edges
of~$\trap_\pi$.  Intuitively, these edges are computed by dragging the
ray through $\pi$ parallel to the $y$-axis until the triangle hit by
the ray changes.  See Figure~2(b).  Let $\triangle_\pi\in\Delta$ be
the triangle visible at~$\pi$, and let $\bar\pi$ be the point on
$\triangle_\pi$ with the same $x$- and $y$-coordinates as $\pi$.  (To
avoid the case where no triangle is visible at $\pi$, we can assume
that there is a large ``background'' triangle.)  Let the
\emph{curtain} of a triangle edge be the set of points on or directly
behind that edge; each curtain is a three-sided unbounded polygonal
slab, two of whose sides are parallel to the
$z$-axis~\cite{bhosk-ershs-94}.  We can find the top (resp.\ bottom)
edge of $\trap_\pi$ by shooting a ray from $\bar\pi$ along the surface
of $\triangle_\pi$ in the positive (resp.\ negative) $y$-direction.
In each case, the desired edge is determined either by an edge of
$\triangle_\pi$ or by the first curtain hit by the ray.  Agarwal and
\Matousek~\cite{am-rsps-93} describe a data structure of size
$O(sn^\e)$, where $s$ can be chosen anywhere between $n$ and $n^2$,
that can answer ray shooting queries in a set of $n$ curtains in time
$O(n^{1+\e}/\sqrt{s})$, after $O(sn^{\e})$ preprocessing time.

\bigskip\noindent\textbf{Stage 3.~Oblique Ray Dragging.}
Each vertical trapezoid edge in $Trap(Vis(\Delta))$ is defined either
by a vertex of $Vis(\Delta)$ at its top or bottom endpoint, or by a
projected visible vertex of some triangle, which could lie anywhere in
the edge.  The third stage looks for the nearest vertices of
$Vis(\Delta)$ along the top and bottom edges of $\trap_\pi$.  Let
$\hat{e}$ and $\check{e}$ be triangle edges whose projections lie
directly above and below $\pi$, respectively, and let
$\hat\pi\in\hat{e}$ and $\check\pi\in\check{e}$ be the points with the
same $x$-coordinate as $\pi$.  Intuitively, we drag rays to the left
and right along $\hat{e}$ (resp.\ $\check{e}$), starting at $\hat\pi$
(resp.\ $\check\pi$), stopping when each ray either hits another edge
or hits an endpoint of $\hat{e}$ (resp.\ $\check{e}$); see
Figure~2(c).  Just as in the previous stage, each ray-dragging queries
can be answered by performing a ray-shooting query in the set of
curtains in time $O(n^{1+\e}/\sqrt{s})$, using Agarwal and \Matousek's
data structure~\cite{am-rsps-93}.

\bigskip\noindent\textbf{Stage 4.~Swath Sweeping.}
In the final stage, we search for the visible triangle vertices whose
projections lie beneath the top edge and above the bottom edge
of~$\trap_\pi$, and whose $x$-coordinates are closest to that of the
pixel~$\pi$.  Since we know that $\triangle_\pi$ is the only triangle
visible in $\trap_\pi$, it suffices to consider only triangle vertices
in front of the plane containing $\triangle_\pi$, and we can assume
that all such vertices are visible.  Intuitively, we take the vertical
swath of rays swept in Stage~2, and sweep it to the left and right
until it hits such a vertex.

We will describe only the leftward sweep; the rightward sweep is
completely symmetric.  It suffices to build a data structure storing
only the rightmost vertex of each triangle, \ie, the vertex with
largest $x$-coordinate.  To answer a swath-sweep query, we perform a
binary search over the $x$-coordinates of the rightmost vertices,
looking for the left edge of $\trap_\pi$.  At each step in the binary
search, we determine whether a particular query trapezoid $\trap$
contains the projection of any visible triangle vertex.  Intuitively,
at each step, we cast a trapezoidal beam forward into the triangles
and ask whether it encounters any triangle vertex before it hits
$\triangle_\pi$.  In fact, since the trapezoid $\trap$ lies entirely
inside the projection of $\triangle_\pi$, it suffices to check whether
the beam hits a vertex before the plane containing $\triangle_\pi$.

We answer this trapezoidal beam query using a \emph{multi-level data
structure}.  Multi-level data structures allow us to decompose
complicated queries into simpler components and devise independent
data structures for each component.  The size (resp.\ query time) of a
multi-level structure is the size (resp.\ query time) of its largest
(resp.\ slowest) component, times an additional factor of~$O(\log n)$
per ``level''.  See \cite{ae-grsir-99,m-rsehc-93} for detailed
descriptions of this standard technique.

We decompose trapezoidal beam queries by observing that the beam 
through a trapezoid $\trap$ contains a visible vertex $v$ if and only 
if
\begin{itemize}\itemsep0pt
\item[(a)]
	the $x$-coordinate of $v$ is between the left and right
	$x$-coordinates of $\trap$,
\item[(b)]
	the $xy$-projection of $v$ is below the top edge of~$\trap$,
\item[(c)]
	the $xy$-projection of $v$ is above the bottom edge of~$\trap$, 
	and
\item[(d)]
	$v$ is in front of the plane containing $\triangle_\pi$.  
\end{itemize}

The first level of our data structure is a range tree~\cite{b-mdc-80}
over the $x$-coordinates of the triangle vertices, which lets us
(implicitly) find the vertices between the left and right sides of 
$\tau$ in $O(\log n)$ time.  This level requires $O(n)$ space and
$O(n\log n)$ preprocessing time.

The next two levels let us (implicitly) find all the vertices whose 
$xy$-projections lie in the wedge determined by the top and bottom 
edges of $\trap$.  One level finds the points below the top edge; the 
other finds the points above the bottom edge.  For each level, we can 
use a two-dimensional halfplane query structure of Agarwal and 
Sharir~\cite{as-anspt-93}, which answers queries in time 
$O((n/\sqrt{s})\log n)$ using space $O(s)$ and preprocessing time 
$O(s\log n)$, for any $s$ between $n$ and~$n^2$.

Finally, in the last level, we need to determine whether any vertex
lies in front of the plane containing $\triangle_\pi$.  We can answer
this three-dimensional halfspace emptiness query in $O(\log n)$ time,
$O(n)$ space, and $O(n\log n)$ preprocessing time using (for example)
a Dobkin-Kirkpatrick hierarchy \cite{dhks-isccd-90}.

Combining all four levels, we obtain a data structure of size
$O(s\log^3 n)$, with preprocessing time $O(s\log^4 n)$, that can
answer any trapezoidal beam query in time $O((n/\sqrt{s})\log^4 n)$,
for any ${n\le s\le n^2}$.  Thus, the overall time to answer a
swath-sweep query is $O((n/\sqrt{s})\log^4 n)$.

\bigskip\noindent Putting all four stages together, we obtain the 
following result.  The time and space bounds are dominated by the 
curtain ray-shooting data structure in the second and third stages.

\begin{lemma}
\label{L:onetrap}
Let $\Delta$ be a set of $n$ disjoint triangles in $\Real^3$, and let 
$s$ be a parameter between $n$ and~$n^2$.  We can build a data 
structure of size $O(sn^\e)$ in time $O(sn^\e)$, so that for any point 
$\pi$ in the $xy$-plane, we can construct the trapezoid $\trap_\pi \in 
Trap(Vis(\Delta))$ containing $\pi$ in time $O(n^{1+\e}/\sqrt{s})$.
\end{lemma}

\section{All Trapezoids}
\label{S:alltraps}

\subsection{Guessing the Output Size}

Lemma~\ref{L:onetrap} implies that for any positive integer~$t$, the
total time to build our data structure and construct $t$ trapezoids is
\[
	O\left(\left(s + \frac{tn}{\sqrt{s}}\right) n^\e\right).
\]
If we know the number of trapezoids in advance, we can minimize the 
total running time by setting $s=\max(n, t^{2/3}n^{2/3})$; the 
resulting time bound is $O(n^{1+\e} + t^{2/3}n^{2/3+\e})$

In our application, however, $t$ is the number of trapezoids in
$\SVM$, which is not known in advance.  We can obtain the same overall
running time in this case using the following standard doubling trick,
previously used in several output-sensitive analytic hidden surface
removal algorithms \cite{os-itosh-94,as-anspt-93,bhosk-ershs-94}.  Our
algorithm runs in several phases.  In the $i$th phase, we build the
data structures from scratch with $s=2^{2i/3}n$, and then construct
the next $2^i\sqrt{n}$ trapezoids.  The time for the $i$th phase is
$O(2^{2i/3}n^{1+\e})$, and the algorithm goes through
$\ceil{\log_2(t/\sqrt{n})}$ phases before it builds all $t$
trapezoids.

\subsection{Avoiding Redundant Queries}

To construct the entire collection of trapezoids $\SVM$, we loop
through the pixels, constructing the trapezoid containing each pixel.
Of course, if we have already built the trapezoid containing a pixel,
we want to avoid building it again.  There are at least two methods
for avoiding this redundancy.

In one method, after we construct each new trapezoid, we search for
and mark all the pixels it contains.  This can be done in
$O((n/\sqrt{s})\log^3 n + k)$ time using a two-dimensional range
searching data structure similar to the one used in the last stage of
our trapezoid-construction algorithm \cite{as-anspt-93}.  Here, $s$ is
as usual an arbitrary parameter between $n$ and $n^2$, and $k$ is the
number of pixels marked.  Since the leading term is dominated by the
time to construct the trapezoid in the first place, this approach adds
only an $O(p)$ term to the overall running time of our hidden-surface
removal algorithm.

\begin{theorem}
Let $\Delta$ be a set of $n$ disjoint triangles in $\Real^3$, and let
$P$ be a set of of $p$ points in the $xy$-plane.  We can construct
$\SVM$ in time $O(n^{1+\e} + t^{2/3}n^{2/3+\e} + p)$, where $t$ is the
number of trapezoids in $\SVM$.
\end{theorem}

Alternately, before querying a new pixel, we could first check whether
it is contained in an earlier trapezoid by performing a point location
query.  We can maintain a semi-dynamic set of $t$ interior-disjoint
vertical trapezoids and answer point-location queries in $O(\log t)$
time per query and $O(\log t)$ amortized time per insertion, using a
data structure of size $O(t\log t)$ based on a segment tree with
fractional cascading \cite{cg-fc1ds-86,cg-fc2a-86,mn-dfc-90}.  This
approach adds $O(p\log t)$ to the overall running time of our
hidden-surface removal algorithm; the total insertion time $O(t\log
t)$ is dominated by other terms.  Although this approach is slower
than pixel-marking, it can be used when the set of pixels is presented
online instead of being fixed in advance.

\begin{theorem}\label{thm.tv}
Let $\Delta$ be a set of $n$ disjoint triangles in $\Real^3$, and let
$P$ be a sequence of $p$ points in the $xy$-plane.  We can maintain
$\SVM$ as points in $P$ are inserted, in total time $O(n^{1+\e} +
t^{2/3}n^{2/3+\e} + p\log t)$, where $t$ is the number of trapezoids
in $\SVM$.
\end{theorem}

\section{A Faster Sweepline Algorithm\\(``Traps and Gaps'')}
\label{S:grid}

The algorithms described in the previous section work for arbitrary 
sets of pixels.  However, in most applications of hidden surface 
removal, the pixels form a regular integer grid.  In this case, we can 
improve the performance of our algorithm using the following 
sweep-line approach, suggested by Pavan Desikan and Sariel 
Har-Peled\cite{dh-pc-99}.

\begin{figure}
\centerline{\footnotesize
\begin{tabular}{c}
	\epsfig{file=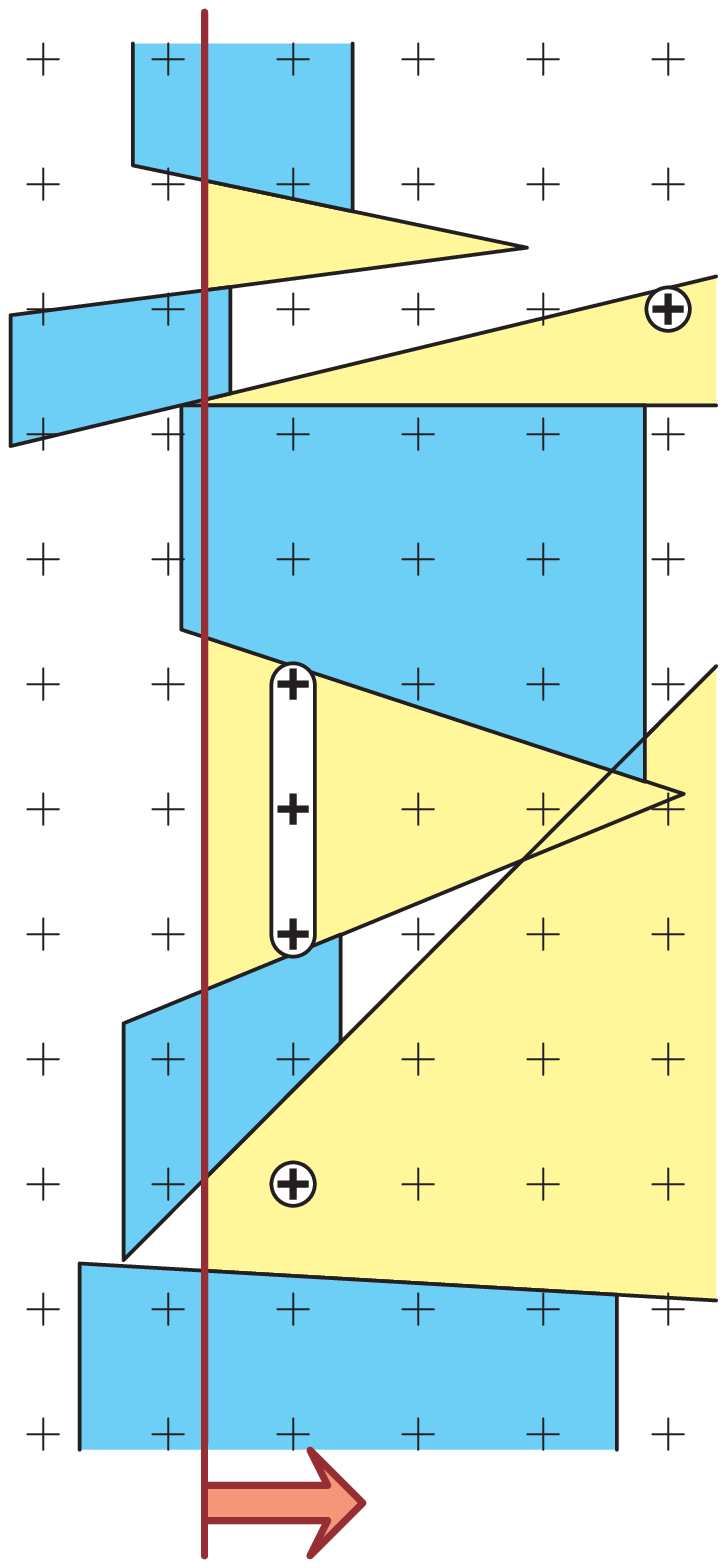,height=2.75in} \\ (a)
\end{tabular}
\hfil\hfil
\begin{tabular}{c}
	\epsfig{file=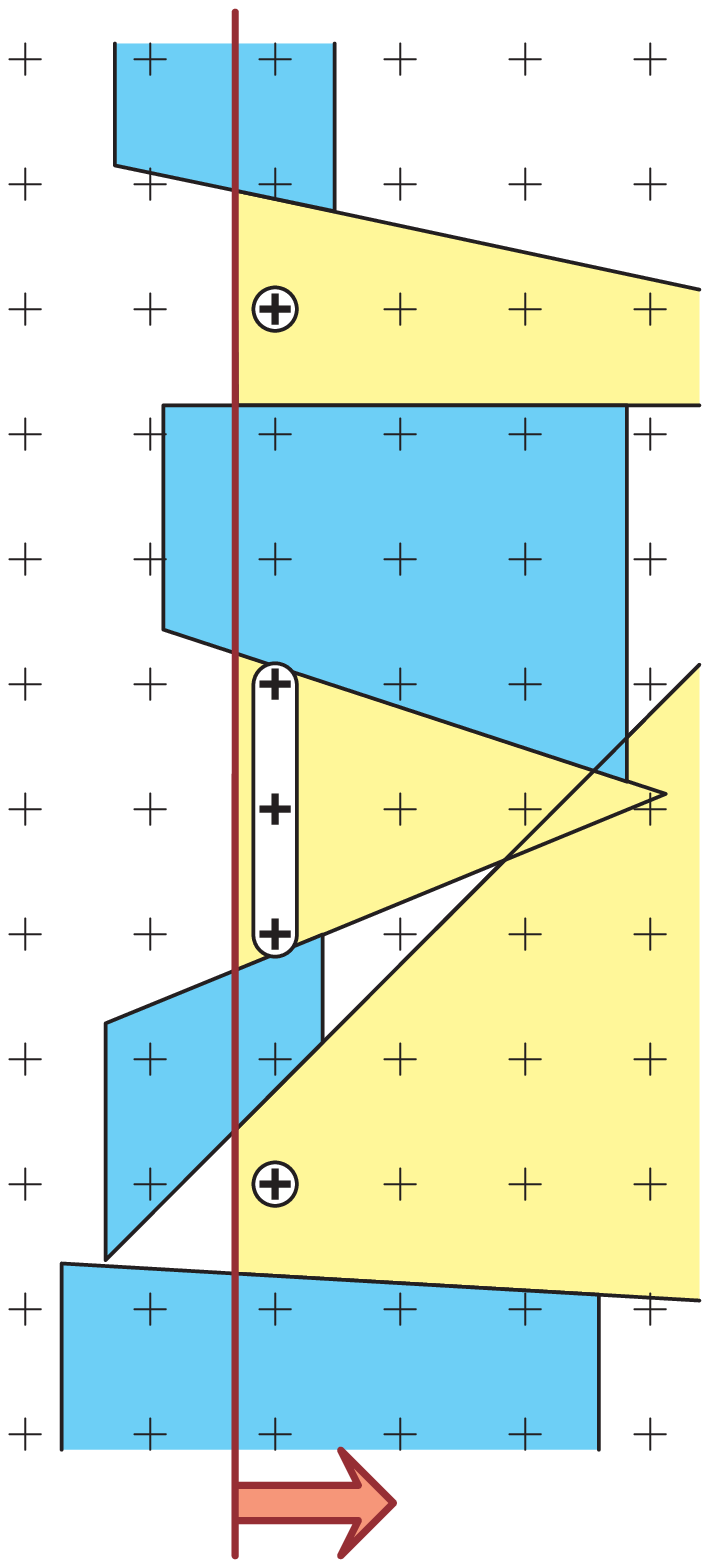,height=2.75in} \\ (b)
\end{tabular}
}
\caption{(a) Just before and (b) just after the sweepline crosses the
right edge of a trapezoid and its neighboring gaps are merged.
Leftmost pixels in each gap, if any, are circled.}
\label{fig:gap}
\end{figure}

Without loss of generality, we assume that the pixel lattice is 
aligned with the coordinate axes.  Our improved algorithm sweeps a 
vertical line $\ell$ across the image plane from left to right.  At 
any position, $\ell$ intersects several trapezoids in $\SVM$.  Between 
any pair of such trapezoids is a \emph{gap}, which is a possibly 
unbounded, possibly empty triangle bounded on the left by $\ell$, 
bounded above by the line through the bottom edge of the higher 
trapezoid, and bounded below by the line though the top edge of the 
lower trapezoid.  Gaps can intersect each other, as well as other 
trapezoids that hit $\ell$.  See Figure \ref{fig:gap}~(a).

We store the traps and gaps in two data structures: a balanced binary 
search tree and a priority queue.  The binary tree stores the traps 
and gaps in sorted order from top to bottom along $\ell$.  For the 
priority queue, the priority of a trap is the $x$-coordinate of its 
right edge, and the priority of a gap is the $x$-coordinate of the 
leftmost pixel(s) inside the gap, or $\infty$ if the gap contains no 
pixels.  Since the sweepline clearly crosses at most $t$ trapezoids, 
the cost of inserting or deleting a trap or gap from the sweep 
structures is $O(\log t)$.  Note that this is bounded by both $O(\log 
p)$ and $O(\log n)$.

To find the leftmost pixel inside a gap, we use the following
two-dimensional integer programming result of
Kanamaru~\etal~\cite{kna-eegpp-94}; see also
\cite{k-patvi-80,f-fatvi-84}.  For related results on enumerating
integer points in convex polygons, see
\cite{lc-avcpg-92,lc-regpc-93,h-osadc-98,h-ctdih-99}.

\begin{lemma}[Kanamaru~\etal~\cite{kna-eegpp-94}]
\label{L:leftmost}
Given a convex $m$-gon $\Pi$, we can find the lowest leftmost integer
point in $\Pi$, or determine that $\Pi$ contains no integer points, in
time $O(m+\log\delta)$, where $\delta$ is the length of the shortest
edge of the axis-aligned bounding box of $\Pi$.
\end{lemma}

\begin{corollary}
We can find a leftmost pixel in any gap, or determine that there is no
such pixel, in $O(\log p)$ time.
\end{corollary}

We do not require that the sweepline structures always contain every 
trapezoid in $\SVM$ that intersects $\ell$.  Instead we maintain the 
following weaker invariant: whenever $\ell$ reaches a pixel $\pi$, the 
trapezoid $\trap_\pi \in \SVM$ containing $\pi$ must be stored in the 
sweepline structures.  We initialize the sweep structure with a single 
gap that contains the entire pixel grid.

When the sweepline $\ell$ reaches the right edge of a trap $\trap$, we
delete it from the sweep structure.  We also delete the gaps
immediately above and below~$\trap$ and insert the new larger gap.
Manipulating the sweep structure requires $O(\log t)$ time, and
finding a leftmost pixel in the new gap requires $O(\log p)$ time, so
the total time required to kill a single trap is $O(\log p)$.

When $\ell$ reaches a leftmost pixel $\pi$ in a gap $\gamma$, we 
perform a trapezoid query to find the trap $\trap_\pi\in\SVM$ 
containing $\pi$.  We then delete $\gamma$ from the sweep structure, 
insert $\trap_\pi$, and insert the two smaller gaps $\gamma^+$ and 
$\gamma^-$ immediately above and below $\trap_\pi$.  The new trap 
$\trap_\pi$ may not contain all the leftmost pixels in $\gamma$; any 
omitted pixels will now be a leftmost pixel in either $\gamma^+$ or 
$\gamma^-$.  If some new gap contains a leftmost pixel of $\gamma$, it 
will be (recursively) filled before the sweepline moves again.  (We 
can avoid creating such ``transient'' gaps by storing the highest and 
lowest leftmost pixels in each gap $\gamma$, at an additional cost of 
$O(1)$ time when $\gamma$ is created, but this improves the running 
time of our algorithm by at most a constant factor.)  For each new 
trap inserted, our algorithm spends $O(\log p)$ time and creates at 
most two new gaps.

Every gap except the initial one is created when a trap is inserted or
deleted.  We can charge at most three gaps to each trap: the gaps
immediately above and below when the trap is inserted, and the gap
left behind when the trap is deleted.  The total number of gaps
created over the entire algorithm is therefore at most $3t+1$.  It
follows that the total time spent finding leftmost pixels is $O(t\log
p)$, and the total time spent manipulating the sweep structures is
$O(t\log t)$.  All the remaining time is spent on trapezoid queries, 
as in our earlier algorithms.

\begin{theorem}
\label{thm:sweep}
Let $\Delta$ be a set of $n$ disjoint triangles in $\Real^3$, and let
$P$ be a regular lattice of $p$ points in the $xy$-plane.  We can
construct $\SVM$ in time $O(n^{1+\e} + t^{2/3}n^{2/3+\e} + t\log p)$,
where $t$ is the number of trapezoids in $\SVM$.
\end{theorem}

Note that this time bound is sublinear in $p$ unless $t=\Omega(p/\log
p)$.  Moreover, the $O(t\log p)$ term is dominated by other terms
unless either $t$ is nearly quadratic in $n$ or $p = 2^{\Omega(n^c)}$
for some positive constant $c$.

\section{Discussion and Open Problems}
\label{S:outro}

One interesting special case of hidden-surface removal is the
so-called \emph{window rendering} problem, where the objects are
axis-aligned horizontal rectangles.  A simple modification of our
algorithm solves this problem in time $O(n\log^2 n + t\log n + p)$
which compares favorably with the best analytic solutions
\cite{b-hsrr-90,gao-osmrh-93}.  If the pixels form a regular grid, we
can improve the running time to $O(n\log^2 n + t\log n)$ using the
sweepline approach.  (Note that this time bound does not depend at all
on the number of pixels!)  Similar improvements can be made for
$c$-oriented polyhedra \cite{bo-hsrco-92}.  It seems likely that our
techniques can also be extended to other special cases of hidden
surface removal with faster analytic solutions, such a polyhedral
terrains \cite{rs-eoshs-88i} and objects whose union has small
complexity \cite{kos-ehsro-92,ho-smhsr-94}.

Perhaps the most interesting open question is whether sampled
visibility maps, or some other similar image decomposition, can be
constructed efficiently \emph{in practice}.  As we mentioned in the
introduction, ray-shooting queries are already answered in practice by
walking through a spatial decomposition defined by the input objects.
The same spatial decomposition can also be used to answer ray-dragging
queries \cite{m-ehsrt-98} and trapezoidal beam queries.  Since curved
models are often polygonalized (and complex polyhedral models are
often simplified) so that each polygonal facet covers only a few
pixels, a practical implementation may require the sampled visibility
map to be redefined in terms of higher-level objects, such as convex
polyhedra or algebraic surface patches, instead of triangles.

A practical implementation of our ideas would have other interesting
applications.  By changing the order in which our algorithm processes
pixels, we can make it suitable for progressive rendering, where the
quality of the image improves smoothly over time as finer and finer
details are computed, or foveated rendering, where fine details are
more important in certain areas of the image than others.  Another
possible application is occlusion culling
\cite{cfhz-cvsov-98,ct-rtocm-97,hmclhz-aocus-97,lg-pmsfe-95,t-vcdop-92}.
By sampling the visibility map at a small number of random points, we
can quickly establish a set of simple occluders that can be used for
conservative visibility tests.  The occlusion tests themselves would
be slightly simpler than in earlier approaches: A triangle is
invisible if its projection is contained in some trapezoid.

Sampled visibility maps exploit spatial coherence well in a global
sense; the number of regions is never much larger than the size of the
visibility map.  In a more local sense, however, there is clearly room
for improvement.  Consider an image that contains mostly empty space,
except for a large number of small triangles near the boundary.
The sampled visibility map consists of several tall thin trapezoids,
but a better decomposition would have a single region covering most of
the image.  It would be interesting to develop decompositions with
better local behavior---perhaps where the expected size of the
component containing a random pixel is maximized, or where the size of
a component is tied to the local feature size \cite{r-draq2-95} of the
visibility map near that component---but with the same global
properties as sampled visibility maps.

\paragraph{Acknowledgment.}
I thank Pavan Desikan and Sariel Har-Peled for suggesting the
sweep-line approach described in Section \ref{S:grid} and pointing out
several relevant references \cite{h-osadc-98,h-ctdih-99,lc-avcpg-92}.

\bibliographystyle{abuser}
\bibliography{gridvis,geom}

\end{document}